\documentclass{article}
\usepackage{multirow}

\usepackage[nonatbib, preprint]{neurips_2024}
\usepackage[numbers]{natbib}
\usepackage[utf8]{inputenc} 
\usepackage[T1]{fontenc}    
\usepackage{hyperref}       
\usepackage{url}            
\usepackage{algorithm}      
\usepackage{algorithmicx}      
\usepackage{algpseudocode}
\usepackage{amsmath}
\usepackage{booktabs}       
\usepackage{amsfonts}       
\usepackage{nicefrac}       
\usepackage{microtype}      
\usepackage{xcolor}         
\usepackage{amsmath}
\usepackage{graphicx}
\usepackage{mathtools, amssymb, lipsum, nccmath} 
\usepackage{placeins}
\usepackage{enumitem} 
\usepackage{siunitx} 

\title{polyGen - A Learning Framework for Atomic-level Polymer Structure Generation}

\author{%
  Ayush Jain\\
  School of Materials Science and Engineering \\ 
  Computational Science and Engineering\\
  Georgia Institute of Technology\\
  \texttt{ayush.jain@gatech.edu} \\ 
  \And
Rampi Ramprasad\\
  School of Materials Science and Engineering\\
  Georgia Institute of Technology\\
  \texttt{rampi.ramprasad@mse.gatech.edu} \\
}

\begin{document}

\maketitle

\begin{abstract}
Synthetic polymeric materials underpin fundamental technologies in the energy, electronics, consumer goods, and medical sectors, yet their development still suffers from prolonged design timelines. Although polymer informatics tools have supported speedup, polymer simulation protocols continue to face significant challenges in the on-demand generation of realistic 3D atomic structures that respect conformational diversity. Generative algorithms for 3D structures of inorganic crystals, bio-polymers, and small molecules exist, but have not addressed synthetic polymers because of challenges in representation and dataset constraints. In this work, we introduce polyGen, the first generative model designed specifically for polymer structures from minimal inputs such as the repeat unit chemistry alone. polyGen combines graph-based encodings with a latent diffusion transformer using positional biased attention for realistic conformation generation. Given the limited dataset of 3,855 DFT-optimized polymer structures, we incorporate joint training with small molecule data to enhance generation quality. We also establish structure matching criteria to benchmark our approach on this novel problem. polyGen overcomes the limitations of traditional crystal structure prediction methods for polymers, successfully generating realistic and diverse linear and branched conformations, with promising performance even on challenging large repeat units. As the first atomic‐level proof‐of‐concept capturing intrinsic polymer flexibility, it marks a new capability in material structure generation.
\end{abstract}

\section{Introduction}

Polymeric materials play a central role in modern science and engineering, enabling technologies across sectors such as packaging, electronics, medicine, and energy. \cite{tran2024nrm, rampiNRM} Their versatility arises from the vast structural diversity of organic building blocks \cite{shukla2023polymer}, the ingenuity of synthetic chemistry, and the breadth of accessible processing techniques. By tuning parameters such as monomer composition, chain architecture, additives, and processing conditions, polymers can be engineered to span a wide range of mechanical, electrical, and transport properties—from rigid plastics, elastomers, dielectrics \cite{gurnani2024ai}, and membranes \cite{phan2024gas, chen2021polymer, PG}. Despite the impact of polymers, the discovery and deployment of new materials remains a slow and resource-intensive process. This is due to the vastness of chemical and processing design spaces \cite{polygnn} and the need to balance performance, cost, safety, and manufacturability. As a result, polymer innovation still relies heavily on experience, trial-and-error, chemical intuition, and serendipity. Novel methods of generating polymer designs guided by informatics approaches have emerged, such as virtual forward synthesis, evolutionary algorithms, and syntax-directed autoencoders that can produce a theoretically innumerable amount of polymer candidates \cite{kern2024informatics, tran2024nrm, kern2021design, batra2020polymers}. Physics-based computer simulations may accelerate the pace of polymer discovery but these methods also face barriers that have thus far prevented the widespread utilization of such approaches as detailed below.

\begin{figure}
    \centering
    \includegraphics[width=0.75\linewidth]{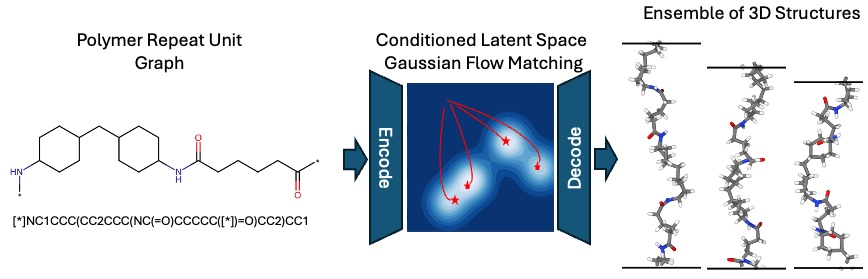}
    \caption{Theoretical overview of polyGen from the perspective of chemistry-conditioned energy minimization of the potential energy surface represented in the latent space. With this capability, an initial connectivity for a polymer repeat unit can be used to generate structures with a high probability of being at a potential energy minimum.}
    \label{fig:intro}
\end{figure}
A key challenge faced by the polymer simulation community is the creation of suitable initial atomic-level structures especially for novel chemistries. Unlike crystalline inorganic materials, polymers exhibit a complex combination of amorphous, semicrystalline, and crystalline domains, with conformational flexibility and structural disorder playing essential roles in their function \cite{huan2020polymer}. Currently, Polymer Structure Predictor (PSP) \cite{sahu2022polymer} is a physics and heuristic-based framework that can generate initial structures of polymers to be used as inputs to Density Functional Theory (DFT) or Molecular Dynamics (MD) simulations. However, rapidly producing statistically diverse and realistic conformations requires accounting for the many low-energy conformations that a polymer can adopt. As illustrated in Figure \ref{fig:intro}, there is a growing need for predictive tools that can generate realistic 3D atomic structures of polymers from minimal inputs, such as chemical composition or connectivity (e.g., SMILES). Such tools would enable property prediction, surrogate simulation, and rational design faster in the discovery pipeline.

\textbf{Generative Materials Structure Prediction} Materials generation via diffusion models \cite{yang2023diffusion} or flow matching \cite{lipman2022flow} offers promise. Typically, these models use datasets containing structures optimized by DFT \cite{tran2023open} or MD simulations \cite{jumper2021highly}. With materials research, these methods have been used thus far for inorganic crystalline materials with a finite number of atoms within a unit cell parameterized by the lattice angles and lengths \cite{jiao2023crystal, xie2021crystal, miller2024flowmm}. Significant progress has also emerged in structure prediction for large biological polymers, particularly proteins, where models like AlphaFold  \cite{jumper2021highly, abramson2024accurate} and Boltz-1 \cite{wohlwend2024boltz} can now generate accurate three-dimensional conformations from amino acid sequences. Similarly, the generation of small molecule conformers \cite{hassan2024flow, xu2023geometriclatentdiffusionmodels, wu2018moleculenet} has also been an open problem, especially in the context of protein docking \cite{corso2022diffdock}.

The synthetic polymers generation problem is a mix of all of these applications. Similar to crystals, synthetic polymers can be defined by a periodic unit. 3D structures of these repeat units are also defined within unit cells/bounding boxes \cite{huan2020polymer}. Traditional crystal structure prediction methods, while powerful for smaller systems, have not displayed abilities to capture the rich conformational landscape of polymers. The local features of synthetic polymers are more akin to molecules where the local (short-range) connectivity of their structures is well-defined. However, polymers occur as long chains composed of sequences of 10s-10,000s of monomeric repeat units, which necessitates an understanding of longer-range interactions for stochastic structure generation. This is similar to protein modeling approaches like AlphaFold3, which employs diffusion transformers to predict an ensemble of structures \cite{abramson2024accurate}. However, unlike proteins, which have a consistent backbone (a sequence of repeating $N-C_{\alpha}-C$, where $C_{\alpha}$ is the centrally located carbon in the amino acid residue) and a finite set of amino acids, synthetic polymers boast a limitless design space to draw their repeat units and backbones from \cite{tran2024nrm, rampiNRM}. Additionally, proteins and molecules are aperiodic and non-periodic, respectively. In combination with the aforementioned challenges, polymer structure datasets from MD or DFT have only recently been standardized and have not seen the scale required for generative modeling \cite{huan2020polymer}. Moreover, past work \cite{wang2025polyconf} in general may overlook the need of representing chain level \cite{huan2020polymer}, amorphous \cite{afzal2020high, phan2024gas}, and network systems \cite{orselly2022molecular}, in simulations using periodic boundary conditions.


\begin{figure}
    \centering
    \includegraphics[width=1.0\linewidth]{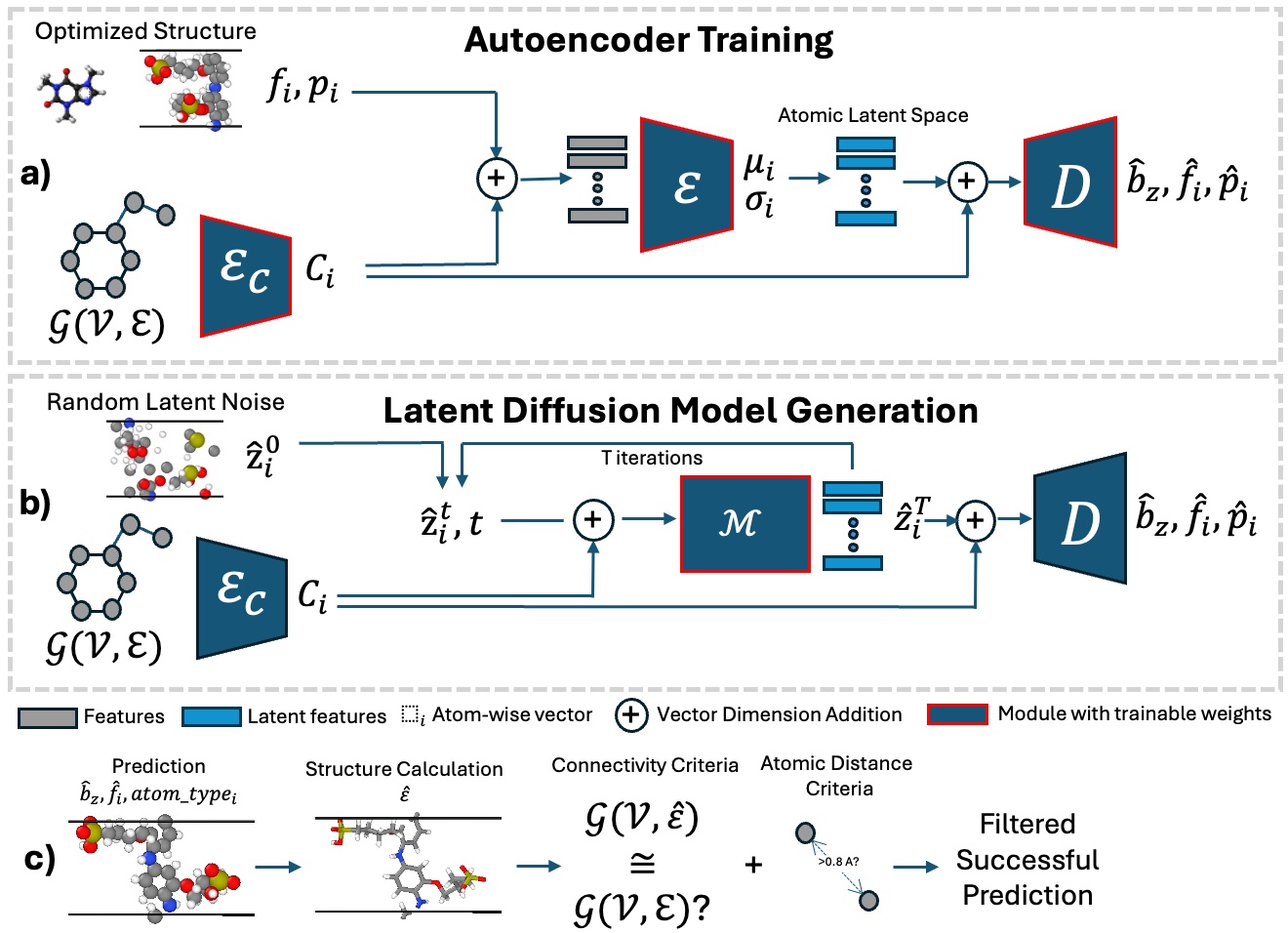}
    \caption{The training and generation process for polyGen. a) First, an autoencoder with conditional encoder $\mathcal{E}_C$, encoder $\mathcal{E}$, and decoder $\mathcal{D}$ is trained to learn an atom-wise latent space by reconstructing the system as a bounding box $\hat{b}_z$, fractional coordinates $f_i$, and cartesian coordinates $p_i$. In this process, an atom-wise conditioning $C_i$ is learned. b) The diffusion model $\mathcal{M}$ is trained within this latent space, iteratively denoising a gaussian random latent $\hat{z}^0_i$ into a new distribution $\hat{z}^T_i$ that is likely a valid polymer conformation. c) During post-generation filtering, a structure is calculated from the predicted positions, and can be used if connectivity and bonding are preserved.}
    \label{fig:arch}
\end{figure}

The complexity of polymer design and representation, combined with limited available data, explains why generative models for polymers have remained largely unexplored until now. We introduce polyGen (Figure \ref{fig:arch}), the first latent diffusion model specifically designed to generate periodic and stochastic polymeric structures. We build upon latent diffusion frameworks for materials generation such as All-Atom DiT \cite{joshi2025all}. Our problem and approach are modified to predict an ensemble of low-energy polymer conformations conditioned on a specified repeat unit with prescribed atomic connectivities rather than performing unconstrained de novo structure generation. Our method leverages a molecular encoding that captures atomic connectivity, which is used as a conditioning signal throughout the model architecture. We adopt a latent diffusion strategy over joint diffusion methods such as DiffCSP \cite{jiao2023crystal} or molecular generation methods in Cartesian space \cite{hassan2024flow} due to the strong coupling between atomic positions and box dimensions imposed by polymer connectivity constraints. We augment our dataset with smaller structures from DFT-optimized molecules, demonstrating improved polymer structure prediction capabilities owing to the shared weights of a dataset invariant encoder and decoder, and a shared latent space. This work culminates in a learning framework that can generate ensembles of realistic polymer chain conformations. To benchmark the quality of these conformations, we introduce rigorous evaluation criteria on bonds, angles, and dihedrals—standards that, to our knowledge, have not yet been applied in the context of materials generation. This application represents a novel proof-of-concept to predict atomic-level synthetic polymer conformations while accounting for their intrinsic flexibility and the distribution of plausible structures.

\section{Results}
\label{sec:results}

\subsection{Dataset and Evaluation Metrics}
\label{sec:dataset}

The main dataset, polyChainStructures, consists of 3855 DFT-optimized infinite polymer chain structures, with a maximum of 208 atoms, including hydrogens. Further details of this already open-sourced dataset, as well as DFT methods can be found in \cite{huan2020polymer}. We use a 3084/386/385 (train/validation/test) split for the polyChainStructures dataset. The QM9 dataset, which contains DFT-optimized small-molecule conformations, is used to augment the training, allowing the models to learn local patterns from molecular structures. We use the train set from QM9 \cite{wu2018moleculenet} which is a size of 100K molecules. We acknowledge the presence of larger molecular datasets (i.e., GEOM) that could be used to further augment our model's learning \cite{axelrod2022geom}. However, from a structural perspective, the conformational behavior of non-zero temperature molecules found in GEOM differs significantly from the idealized infinite chains at 0K modeled in this work.

We benchmark polymer structure prediction on our polyChainStructures test set with our method. Unlike crystals, polymers are amorphous and can adopt many valid low-energy conformations, making conventional structure matching \cite{ong2013python} unsuitable. Molecule generation is also evaluated with root mean squared distance after rotational alignment \cite{xu2023geometriclatentdiffusionmodels, jing2022torsional}, but this doesn't account for conformational variability experienced by infinite chains.

To this end, we compare the predicted conformations against the DFT-optimized structures by examining distributions over bond lengths, bond angles, and dihedrals. We quantify similarity using the forward Kullback–Leibler (KL) divergence from the predicted distribution to the ground truth. The task is to generate a distribution of plausible polymer conformations and assess the likelihood that the DFT-optimized structure could have been drawn from this predicted distribution. For a set of predicted quantities $Q_{\text{pred}}$ we match it to the set of DFT predicted quantities $Q_{\text{DFT}}$ with

\begin{equation}
\label{eq:for-kl}
\mathcal{L}_{\text{KL}} = D_{\text{KL}} \left( Q_{\text{DFT}} \,\|\, Q_{\text{pred}} \right) = \int Q_{\text{DFT}}(z) \log \frac{Q_{\text{DFT}}(z)}{Q_{\text{pred}}(z)} \, dz
\end{equation}

In practice, we do this over a discrete set of buckets where $dz$ is 0.001 of the predefined ranges of the quantities, which could be bond lengths (0.9 to 2.0), angles (0 to 180), or dihedrals (0 to 360).

\subsection{Structure Matching Results}

\begin{figure}[h]
  \centering
  \includegraphics[width=\linewidth]{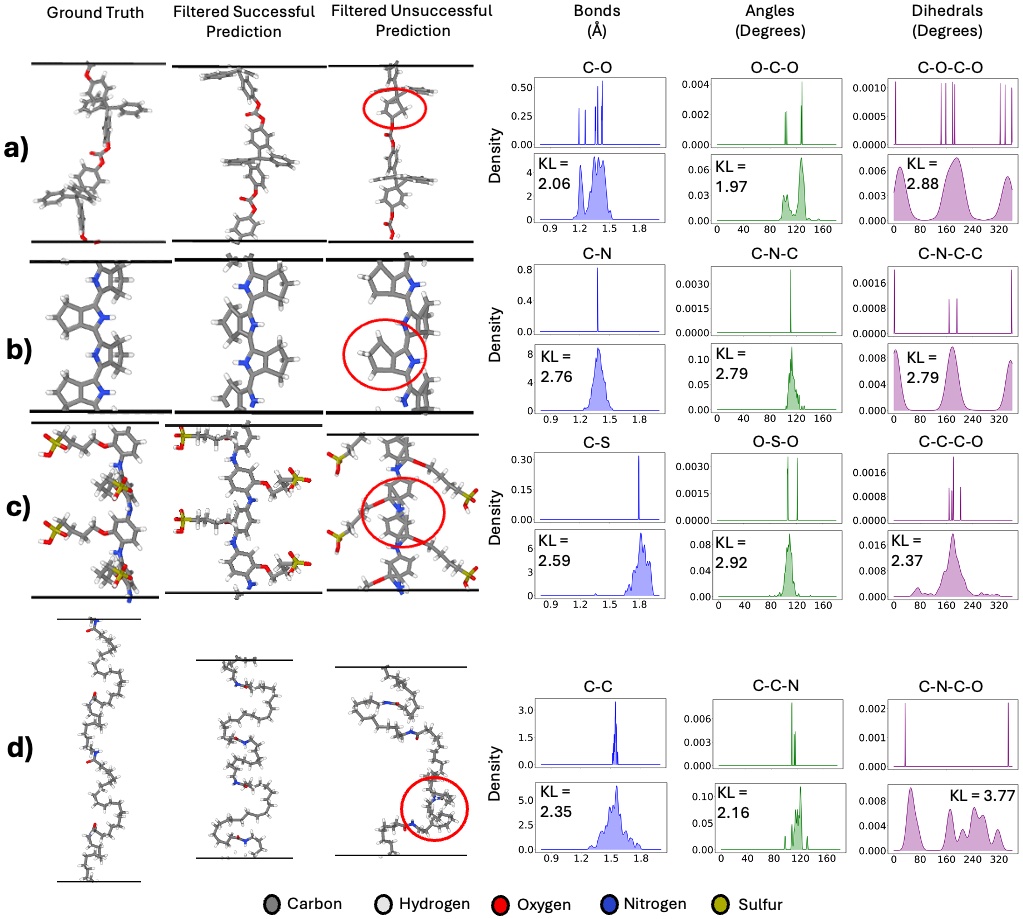} 
  \caption{Visual samples from 100 generations per polymer type. Each row (a-d) displays examples from a single polymer. The ground truth, a successfully generated sample, and an unsuccessful sample are provided. The unsuccessful samples highlight the errors that caused the incorrect connectivity. Note that the structures for b) and c) are replicated along the z-axis for visual clarity of the structure. The visualizations are accompanied by corresponding ground truth distributions (top) and model-predicted distributions (bottom) of molecular bonds, angles, and dihedrals for that polymer.}
  \label{fig:generations}
\end{figure}

We find that our model is capable of predicting polymer structures that closely match the linear chain structure calculated by DFT. To provide context to our rationale, Figure \ref{fig:generations} shows examples of predicted polymers compared to the ground truth of the dataset. In Figure \ref{fig:generations}a-d we see the capabilities of generating systems with qualitatively similar structures after filtering. We also provide examples of errors that are caught by our filtering. In Figure \ref{fig:generations}a and c, we see examples of carbon atoms incorrectly predicted within a supposedly aromatic ring in the backbone, which lead to incorrect connectivity. Many failed structures, especially the one in Figure \ref{fig:generations}b, can easily be fixed with an energy minimization or a heuristic-based increasing of the C-H bond, demonstrating the efficacy of the remaining generation and the stringent nature of the generation filter. Also, the complexity of these structures should be noted, given chain size, branching, and the number of rings present. 

\FloatBarrier

To quantify structural similarity, we compare bond length, triplet angle, and dihedral angle distributions for some atomic species. Predicted structures should exhibit peaks comparable to optimized structures. The forward KL divergences establish benchmarks for acceptable "matches" in bond lengths, angles, and dihedrals. 

We find that our approach can find relative trends in the majority of bond lengths, but lacks precision. For example, the predicted C-O bonds in Figure \ref{fig:generations}a  shows 2 peaks for single and double bonds, with noisy predictions scattered by at least 0.1 {\AA} around these peaks. In Figure \ref{fig:generations}b and c, both predicted C-N and C-S bond distributions show peaks corresponding with the true bond length, but with wide distributions. Generation is precise to the order of {\AA}, but not on the scale of picometers. Precision on the level of picometers will be necessary to properly distinguish between bond types before our approach is scaled to larger system sizes. 

We find that the distribution peaks of angles and dihedrals match the ground truth peaks, especially the O-C-O and O-S-O bonds in Figure \ref{fig:generations}a and c, respectively. The dihedral distributions can be used to validate the prediction of correct local structures in the polymer. For example, the C-N-C-C dihedrals in Figure \ref{fig:generations}b validate the feasible prediction of the nitrogen containing 5-membered ring in the backbone. In Figure \ref{fig:generations}c, the C-C-C-O dihedral, located on a branch, shares a peak with the ground truth at \SI{160}{\degree} but has a few predictions around \SI{80}{\degree} and \SI{320}{\degree}. This reflects the difficulties in generating branched structures, which may have conformational variability.

Figure \ref{fig:generations}d shows predictions for the largest system in the test dataset, containing 208 atoms. Only 3 out of 100 generated structures pass the filtering when using position-biased attention, while the DiT model with vanilla attention fails to produce any valid structure. The broad distribution of predicted C–N–C–O dihedral angles and their high deviation from the ground truth further highlight the lack of conformational viability. These results underscore a clear limitation in handling larger systems, likely due to constraints in the training dataset, and illustrate the challenges of generating feasible structures for complex polymer repeat units.

\begin{figure}[h!]
  \centering
  \includegraphics[width=0.9\linewidth]{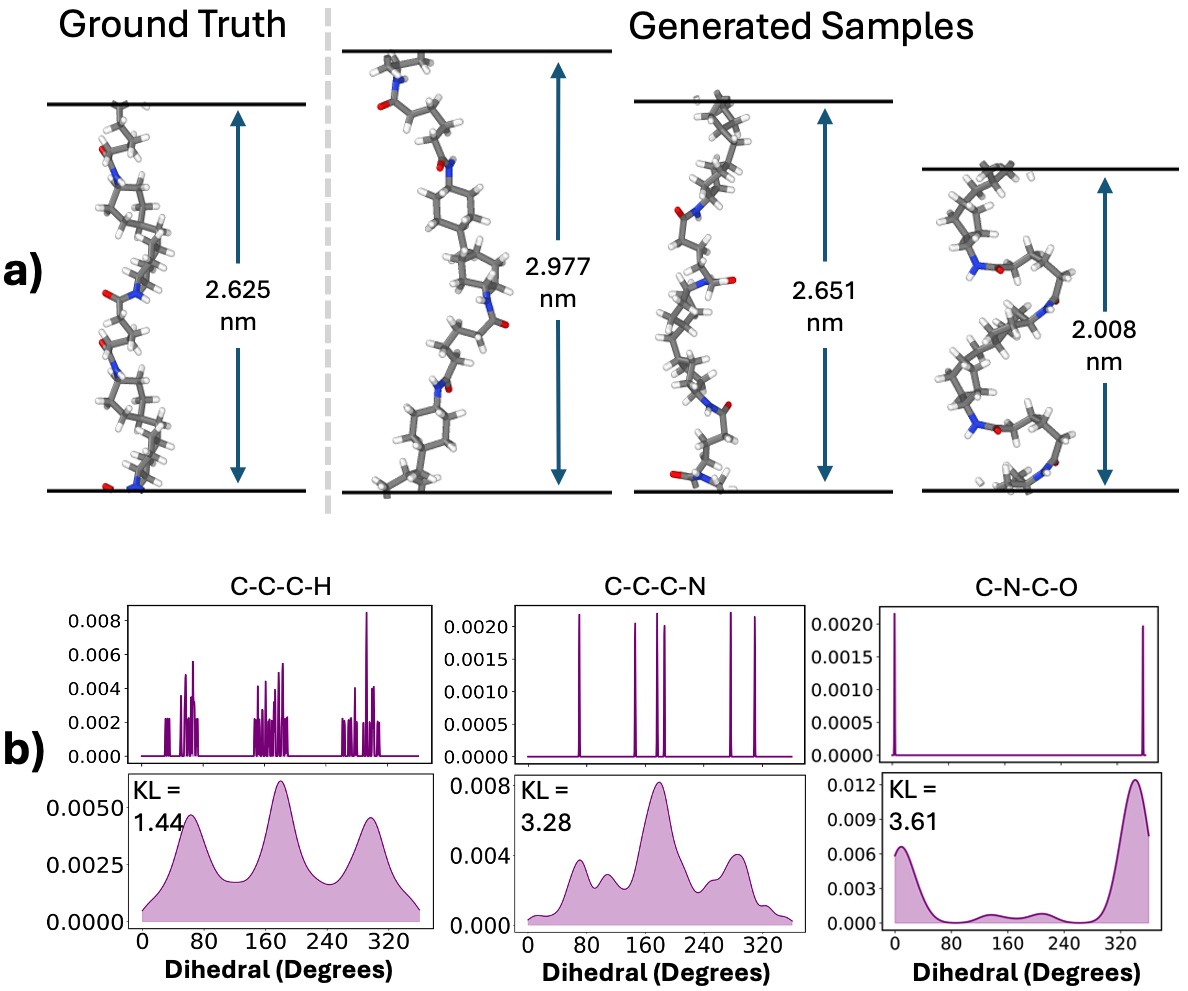} 
  \caption{a) Demonstration of diverse structures generated from polyGen for the same repeat unit, compared to the ground truth structure. b) The ground truth dihedrals (top) compared to the distributions of the full generated ensemble (bottom) along with the KL Divergence. For reference, the structures are generated from the polymer repeat unit in Figure \ref{fig:intro}.}
  \label{fig:diversegen}
\end{figure}

\subsection{Prediction of Diverse Samples}
\label{sec:diverse}

As highlighted earlier, polymer structures do not reside in one fixed conformation but occur as an ensemble of minimized conformations, even close to 0 K. Our training dataset contains only 1 conformational example out of the potential ones, because the generation of these ensembles is costly. A useful generation model would produce a diverse ensemble of conformations representative of many possible low-energy states. In Figure \ref{fig:diversegen}b, despite the lack of per-polymer distribution in our dataset, polyGen can generate an ensemble of diverse structures with variable repeat unit lengths. Moreover, these samples pass the post-generation filtering and remain consistent with structural features in the DFT ground truth, as seen in the KL Divergences in Figure \ref{fig:diversegen}b.

\subsection{Overall Results}

\begin{figure}[b]
  \centering
  \includegraphics[width=0.9\linewidth]{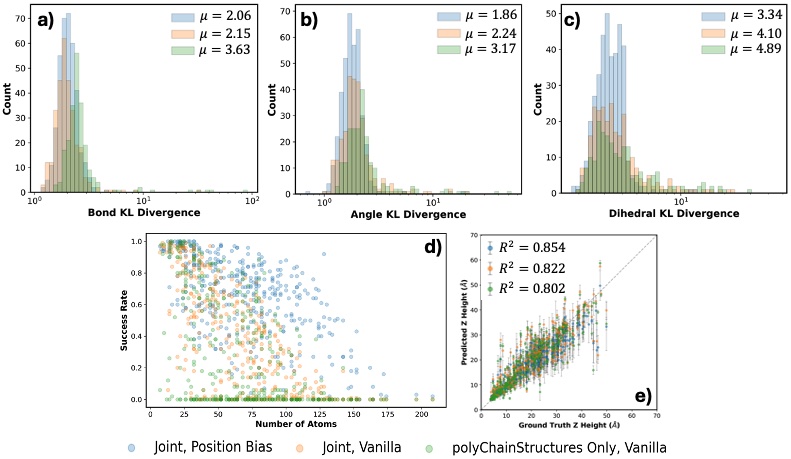} 
  \caption{KL Divergences of Joint training for Position Bias, Joint Training and polyChainStructures Only for a) Bonds b) Angles and c) Dihedrals. d) The comparison of z-height prediction vs the DFT ground truth. e) The success rate of filtering as a function of the atom count of the system.}
  \label{fig:overall_results}
\end{figure}

Following previous works, we also see that joint training on multiple DFT datasets can improve model performance \cite{joshi2025all, tran2023open}. In this section, we show the specific areas in which the inclusion of the QM9 dataset can benefit the generation of polymer structures by comparing the joint dataset approach with a model fully trained on just the polyChainStructures dataset. Figure \ref{fig:overall_results} shows the overall KL divergences for bond lengths, angle lengths, and dihedrals, and comparison of predicted z-height. The inclusion of the QM9 dataset improves predictions on the local levels of polymer chains, as the KL divergences of bonds and angles decrease by 41.0\% and 29.0\%, respectively, upon the addition of the QM9 dataset. Larger features, such as dihedral angles and the z-height show limited improvement, with dihedrals improving by 16.1\%. This is likely because small molecules do not provide any insight into the macrostructural properties of polymers. Overall, we find that 36.1\% of the generated structures by the jointly trained model pass our filtering, whereas 27.4\% samples are generated correctly when using only the polyChainStructures dataset for training.

Another major improvement is seen when including position attention biasing (Section \ref{sec:relbias_method}) in the DiT module. Because local interactions are explicitly specified, we find that the positional encoding improves generation success from 36.9\% to 64.8\% when looking at isomorphism of the predicted structure compared to the original graph. Figure \ref{fig:overall_results} showcases the improvement of polyGen with a relative encoding bias on all metrics, notably the decrease of angle and dihedral KL Divergences by 33.5\% and 21.2\%, respectively. 

As highlighted in Figure \ref{fig:overall_results}d and discussed in Section \ref{sec:diverse}, we expect some variability in the prediction of the z-height of the polymer chain. Despite this, we can achieve an $r^2$ score of 0.854 between the predicted and DFT heights, meaning that the model can differentiate between dense vs. sparsely packed structures. Additionally, we see a larger standard deviation and error in predictions for larger repeat units (> 10 {\AA}), because these repeat units may show more flexibility. Figure \ref{fig:overall_results}e also highlights the difficulty of generation as system size increases. Most generated systems with <25 atoms show success rates > 0.5. For system sizes greater than 150 atoms, the model with vanilla attention cannot generate any systems with proper connectivity, but the relative position bias substantially improves this.

Intuitively, it would also seem that complex polymeric structures would be more difficult to generate. To this end, we also compare these metrics with the Synthetic Accessibility (SA) in Supplementary Information A. We see loose correlations with KL divergences of bond length, angles, and dihedrals with the SA score, but the ratio of successful generations of a polymer do not have a strong correlation with the SA score. Therefore, generation success could be independent from SA score and more reliant on the system size. 

\section{Discussion}

In this study, we introduce a solution to the problem of atomic-level polymer structure generation—given only the atomic connectivity of a repeat unit (i.e. SMILES), polyGen can generate an ensemble of realistic 3D structures of synthetic polymers. Results demonstrate that polyGen effectively generates structures with bond lengths, angles, and dihedral distributions that align well with ground truth structures, and the quality of structures improves with the inclusion of positional biasing to the attention mechanism of the diffusion transformer. The model also successfully generates valid conformations for complex features like aromatic backbones and branches while correlating repeat unit chemistry with structural properties, such as repeat unit length. These initial results are particularly promising considering our polymer structure dataset optimized with Density Functional Theory contains only 3855 systems. We also investigate the limitations and demonstrate the need for stringent benchmarking of future polymer structure generation techniques. While capturing relative trends in structure, the model lacks precision at the picometer scale needed to definitively distinguish between bond types. Performance degrades significantly for larger polymeric systems, likely due to dataset constraints. Finally, the consistency of feasible generation needs to be improved. 
Given the successes of this proof-of-concept, future work will focus on expanding the training dataset to include larger polymer systems, incorporating additional physics-informed constraints, and exploring hybrid approaches that combine latent diffusion with molecular dynamics simulations. Addressing these challenges could transform polyGen into an invaluable tool for computational materials science, accelerating the discovery of novel polymeric materials across numerous applications.

\section{Methods}

polyGen consists of three phases, a 0D conditioning on the molecular graph of the desired polymer repeat unit chemical structure, a variational autoencoder for structure (Figure \ref{fig:arch}a), and a latent diffusion module (Figure \ref{fig:arch}b). Overall, we choose an architecture that does not include any equivariance or inductive biases, following the sentiment of recent works \cite{joshi2025all, abramson2024accurate} where data scale and augmentation can be a path to learning these more efficiently.

\subsection{Polymer Structure Specification}

Several previous works have learned structural and chemical representations of periodic materials, which include fractional coordinates, unit cell lengths, and lattice angles. For single polymer chain structures, we take a modified approach. Because of their complexity, training set structures are optimized using DFT with a periodic orthogonal box where the chain is continuous through the $z$ axis. The width of the box is needed to encapsulate side chains/branched structures and are made large enough to isolate the single chain during optimization.

We note that the $x$ and $y$ axis are not entirely polymer structure dependent. Including these as prediction variables is an overparameterization, and does not contribute much information. Therefore, we fix the $x$ and $y$ of the box to a 55 {\AA} x 55 {\AA} box, and define the system by the $z$ height of the box, $b_z$. In theory, this quantity is a proxy for the system's density (which is also stochastic, depending on the specific sampled configurations). Atomic positions are provided as fractional coordinates within this orthogonal bounding box.

\subsection{Graph Conditioning}

The generation of polymer structures begins with a minimal representation of atomic connectivity, typically provided in the form of a SMILES string. This string is converted into a molecular graph: periodic for polymers and non-periodic for small molecules. To extract a universal atom-wise representation, denoted as \(C_i\), we apply a graph conditioning module that encodes the molecular connectivity. As illustrated in Figure \ref{fig:arch}a and b, \(C_i\) serves as contextual input to the encoder, decoder, and diffusion modules. The graph conditioning module is implemented using a graph interaction network, similar to \cite{battaglia2016interaction}, which captures local chemical environments by modeling interactions between different atom and bond types. Importantly, the same network weights are used across all datasets and connectivity types, facilitating effective transfer learning and generalization across chemical spaces.

Given a graph $\mathcal{G} = (\mathcal{V}, \mathcal{E})$, each node $i \in \mathcal{V}$ is associated with an atom type and positional encodings, and each edge $(i, j) \in \mathcal{E}$ is associated with a bond type. The initial node and edge embeddings are computed as:

\begin{equation}
\begin{aligned}
\mathbf{h}_i^{(0)} &= \text{Embed}(\text{atom\_type}_i), \quad 
\mathbf{e}_{ij}^{(0)} = \text{Embed}(\text{bond\_type}_{ij}), \quad
\mathbf{h}_i^{(0)} = \text{MLP}([\mathbf{h}_i^{(0)}; \mathbf{r}_i^{\text{RW}}; \mathbf{r}_i^{\text{Lap}}])
\end{aligned}
\end{equation}
For each layer $l = 1, \ldots, L$, we update the edges and the nodes as:

\begin{equation}
\begin{aligned}
\mathbf{e}_{ij}^{(l)} &= \mathbf{e}_{ij}^{(l-1)}\mathbf{W}_e + f_e^{(l)}(\text{LN}([\mathbf{e}_{ij}^{(l-1)}; \mathbf{h}_i^{(l-1)}; \mathbf{h}_j^{(l-1)}])) \\
\mathbf{m}_i^{(l)} &= \sum_{j \in \mathcal{N}(i)} \mathbf{e}_{ij}^{(l)}, \quad
\mathbf{h}_i^{(l)} = \mathbf{h}_i^{(l-1)}\mathbf{W}_v + f_v^{(l)}(\text{LN}([\mathbf{m}_i^{(l)}; \mathbf{h}_i^{(l-1)}])) \\
\end{aligned}
\end{equation}

We finally node features to add global context for the whole molecule:
\begin{equation}
\begin{aligned}
\mathbf{g} &= \sum_{i \in \mathcal{V}} \mathbf{h}_i^{(L)}, \quad
C_i = f_f^{(l)}([\mathbf{g}; \mathbf{h}_i^{(L)}])
\end{aligned}
\end{equation}

where $f_e^{(l)}$,  $f_v^{(l)}$, $f_f^{(l)}$ are MLPs, $\text{LN}$ is layer normalization, $\mathbf{r}_i^{\text{RW}}$ is a random walk positional encoding \cite{dwivedi2021graph} of size 16 and $\mathbf{r}_i^{\text{Lap}}$ is a Laplacian positional encoding \cite{dwivedi2023benchmarking, Wang2023SwallowingTB} of size 2. We use $L=4$ to capture local interactions of an atom 1 "hop" away from its furthest dihedral. The global pooling prior to embedding into token dimension is done so that atomic-level information is taken with global system information.

The encoder, decoder, and diffusion modules are all transformers, which typically require a positional encoding to identify token ordering and condition the self-attention weights. Similar to contemporary graph transformer methods, $C_i$ communicates this by providing embeddings calculated from $\mathbf{r}_i^{\text{RW}}$ and $\mathbf{r}_i^{\text{Lap}}$ \cite{min2022transformer}.

\subsection{Structural Variational AutoEncoder}

The next step is to learn a way to combine all aspects of a polymer system, such as monomer chemistry, fractional coordinates, and the bounding box, into a joint space. This space should also allow the fusion of information from small molecules or other systems to enhance learning. We employ a VAE to create unified atom-wise latent representation that contains structural details for small molecules and polymers, $\mathcal{Z} \in \mathcal{R}^d$ where $d$ is a latent dimension. The VAE is based on traditional graph VAEs, including an encoder $\mathcal{E}$ and a decoder $\mathcal{D}$ operating atom-wise.

The structure of the molecule/polymer is given through a concatenated vector of fractional coordinates $f_i$ and Cartesian positions $p_i$. For a polymer structure, both the $f_i$ and scaled $p_i$ according to the bounding box are provided. This forces the VAE to learn and reconstruct the relationship between the Cartesian positions, fractional coordinates, and the bounding box. For non-periodic molecules, $p_i$ is provided and $f_i = \emptyset$, which allows periodic and non-periodic materials to share the same latent space, similar to \cite{joshi2025all}.  The atom-wise latent space is then calculated along with the $C_i$. 

\begin{align}
    \mu_i, \sigma_i = \mathcal{E}(f_i, p_i, C_i) \\
    \mathcal{Z}_i \sim \mathcal{N}(\mu_i, \sigma_i)
    \label{equ:encoding}
\end{align}

The decoder then produces a structure from $\mathcal{Z}$,

\begin{align}
    \hat{f}_i, \hat{p}_i, {\hat{b}}_z = \mathcal{D}(\mathcal{Z}, C_i)
\end{align}

where $\hat{f}_i$ is the predicted fractional coordinates, $\hat{p}_i$ is the predicted cartesian coordinates, and ${\hat{b}}_z$ is the $z$ direction height of the bounding box. The loss on the decoder is used to optimize the model

\begin{equation}
\begin{split}
\mathcal{L}_{\text{total}} &= \langle w_{\text{bbox}} \cdot \mathcal{L}_{\text{bbox}} \rangle + \langle w_{\text{frac\_coords}} \cdot \mathcal{L}_{\text{frac\_coords}} \rangle + \langle w_{\text{pos}} \cdot \mathcal{L}_{\text{pos}} \rangle \\
&+ \langle w_{\text{kl}} \cdot \mathcal{L}_{\text{kl}} \rangle + \langle w_{\text{bond}} \cdot \mathcal{L}_{\text{bond}} \rangle + \langle w_{\text{angle}} \cdot \mathcal{L}_{\text{angle}} \rangle + \langle w_{\text{dihedral}} \cdot \mathcal{L}_{\text{dihedral}} \rangle
\end{split}
\end{equation}

\text{Where the reconstruction terms are defined as:}
\[
\begin{aligned}
\mathcal{L}_{\mathrm{bbox}} &= \mathrm{MSE}\bigl(\hat b_z,\;b_z/(10\sqrt[3]{N})\bigr), &
\mathcal{L}_{\mathrm{frac\_coords}} &= \frac{1}{d}\sum_i \mathrm{MSE}(\hat f_i, f_i),\\
\mathcal{L}_{\mathrm{pos}}  &= \frac{1}{d}\sum_i \mathrm{MSE}\bigl(\hat p_i - \langle \hat p\rangle,\;p_i - \langle p\rangle\bigr), &
\mathcal{L}_{\mathrm{kl}}   &= \mathrm{KL}\bigl(q_{\phi}(z\mid x)\;\|\;p(z)\bigr).
\end{aligned}
\]
\text{And the structural loss terms are:}
\begin{align*}
\mathcal{L}_{\text{bond}} &= \text{MSE}(d_{ij}, \hat{d}_{ij}) & 
\mathcal{L}_{\text{angle}} &= \text{MSE}(\angle_{ijk}, \hat{\angle}_{ijk}) &
\mathcal{L}_{\text{dihedral}} &= (\Delta_{\text{periodic}}(\tau_{ijkl}, \hat{\tau}_{ijkl}))^2
\end{align*}
\text{Where:}
\begin{align*}
d_{ij}, \hat{d}_{ij} &= \text{bond length between atoms $i$ and $j$ with true/predicted coordinates} \\
\angle_{ijk}, \hat{\angle}_{ijk} &= \text{angle between bonded atoms $i$-$j$-$k$ with true/predicted coordinates} \\
\tau_{ijkl}, \hat{\tau}_{ijkl} &= \text{dihedral angle of atoms $i$-$j$-$k$-$l$ with true/predicted coordinates} \\
\Delta_{\text{periodic}} &= \text{periodic difference within 0-2$\pi$ radians}, \; N = \text{number of atoms}
\end{align*}

The inclusion of the structural loss is to help the VAE optimize to decode structures that are structurally similar to the target, even if the positions may be slightly different than the target. Similar approaches for enforcing local atomic relations are utilized in works concerning biological polymer structures \cite{abramson2024accurate, mariani2013lddt}. Final model hyperparameters are provided in Supplementary Information C.

\subsection{Diffusion Transformer}

Now that a joint latent space is learned, we require a way to find suitable structures for new polymer chemistries. We use a DiT architecture for our generative model $\mathcal{M}$, operating within the latent space $\mathcal{Z}$ learned by the VAE. We use a similar DiT architecture as ADiT \cite{joshi2025all}, however the previous work uses sinusoidal positional encodings for atomic tokens, which make it difficult for unordered atomic representations. In our case, we want to preserve the permutation invariant qualities of a GNN, so we use the positional encoded features and interactions in \(C_i\) to provide positional information relative to other atoms in the system. We also modify the attention mechanism to add a bias towards bonds, elaborated in Section \ref{sec:relbias_method}.

Our denoiser is implemented through a gaussian flow matching approach, which is equivalent to denoising diffusion as one can be derived from the other \cite{gao2025diffusionmeetsflow, joshi2025all}. We start by encoding a DFT optimized structure into the latent space, using Equation \ref{equ:encoding}, to get $\mathcal{Z}_1$. Similar to other latent diffusion/flow matching works we denoise from zero-centered random noise $\mathcal{Z}_0 \sim \mathcal{N}(0, 1)$ at $t=0$ to $\mathcal{Z}_1$ at $t=1$ \cite{joshi2025all}. To train the transformer, we provide it with an interpolated sample $\mathcal{Z}_t$ at a random timestep $t \sim \mathcal{U}(0, 1)$,

\begin{equation}
\mathcal{Z}_t = (1 - t)\mathcal{Z}_0 + t{\mathcal{Z}_1}
\end{equation}

We can pose the learning problem as the linear ordinary differential equation (ODE),

\begin{equation}
u_t = \dot{\mathcal{Z}}(\mathcal{Z}_t) = \frac{\mathcal{Z}_1 - \mathcal{Z}_t}{1 - t}
\end{equation}

The final prediction task is defined as,
\begin{align*}
\hat{\mathcal{Z}}_1 = \mathcal{M}(\mathcal{Z}_t, \mathcal{Z}_{\text{sc}}, t, C_i, S), \quad \hat{u}_t = \frac{\hat{\mathcal{Z}}_1 - \mathcal{Z}_t}{1 - t}
\end{align*}

where $S$ is an embedding that represents the dataset used during generation. This is used exclusively during joint training. Only the embedding for the polyChainStructures dataset is used during inference. \(\mathcal{Z}_{\text{sc}}\) denotes a self-conditioning input, which corresponds to a previous prediction of \(\hat{\mathcal{Z}}_1\). Self-conditioning improves autoregressive molecular generation \cite{stark2023harmonic, joshi2025all}. Our training uses a two-pass approach: first predicting $\hat{\mathcal{Z}}_1$ with $\mathcal{Z}_{\text{sc}} = \emptyset$, then feeding this prediction back as $\mathcal{Z}_{\text{sc}}$ for refinement. To avoid overreliance, we randomly drop self-conditioning ($\mathcal{Z}_{\text{sc}} = \emptyset$) with probability 0.5 during training. Within $\mathcal{M}$, the latent features and conditioning signals are incorporated as 

\[
\begin{aligned}
\tilde x=E_{\mathcal{Z}}([\mathcal{Z}_t;\mathcal{Z}_{\text{sc}}]) + C_i, \quad c=d+E_t(t),
\end{aligned}
\]

where $\tilde x$ is input and $c$ is the modulation conditioning for attention blocks from \cite{Peebles2022DiT}. $E_{\mathcal{Z}}$ and $E_t$ are embedding blocks.

The training objective is defined as the atom-wise mean squared error between $\hat{u_t}^{(i)}$ and $u_t^{(i)}$ for $N$ atoms in a system:

\begin{align}
\mathcal{L}_{\mathcal{G}} 
&= \frac{1}{N} \sum_{i=1}^N \left\| u_t^{(i)} - \hat{u}_t^{(i)} \right\|^2 \\
&= \frac{1}{N} \sum_{i=1}^N \left\| \frac{\mathcal{Z}_1^{(i)} - \mathcal{Z}_t^{(i)}}{1 - t} - \frac{\hat{\mathcal{Z}}_1^{(i)} - \mathcal{Z}_t^{(i)}}{1 - t} \right\|^2 \\
&= \frac{1}{(1 - t)^2} \cdot \frac{1}{N} \sum_{i=1}^N \left\| \mathcal{Z}_1^{(i)} - \hat{\mathcal{Z}}_1^{(i)} \right\|^2
\end{align}

To prevent numerical instability in the loss function calculation, we clip the value of $t$ to 0.9, in accordance with previous work \cite{joshi2025all}.

During inference, we sample an initial latent \(\mathcal{Z}_0 \sim \mathcal{N}(0, I)\) and iteratively denoise it using \(T\) steps of Euler integration:
\[
\mathcal{Z}_{t + \Delta t} = \mathcal{Z}_t + \Delta t \hat{u}_t,
\]
where $\Delta t = 1.0/T$. This process generates a final conformation \(\hat{\mathcal{Z}}_1\), which is decoded into a novel 3D atomic system using $\mathcal{D}$. We compare generation hyperparameters in Supplementary Information B, and show final model hyperparameters in Supplementary Information C.

\subsubsection{Relative Positional Encoding Attention Modification}
\label{sec:relbias_method}

With a traditional attention mechanism, DiT is forced to learn bonding relationships from the atom-wise conditional embeddings. For smaller systems, this can provide enough information for proper structure prediction, but this may be a problem for larger systems. Alphafold3, which operates on larger atomic systems than traditional material structure generation models, utilizes a pairformer \cite{abramson2024accurate} with a relative position encoding to allow tokens to attend to nearby neighbors by conditioning the attention weights of the diffusion module.

We employ a lightweight relative positional encoding with attention biasing mechanism in the polyGen DiT module to differentiate local atomic interactions from global. First, we encode pairwise relationships with a one-hot graph-distance tensor
\[
D \;\in\;\{0,1\}^{N\times N\times 5},
\]
where each of the five channels indicates whether atoms \(i\) and \(j\) are
\begin{enumerate}[label=(\arabic*)]
  \item identical
  \item bonded
  \item separated by an angle
  \item separated by a dihedral
  \item beyond four bonds apart
\end{enumerate}
An MLP \(f_{D}\) then maps each one-hot vector \(D_{i,j}\) to a scalar bias, which is added directly into the scaled dot-product attention:

\begin{align}
\mathrm{bias}_{i,j} &= f_{D}\bigl(D_{i,j}\bigr),\\
\mathrm{Attention}_{i,j} &= \mathrm{Softmax}\!\Bigl(
  \frac{Q_i \cdot K_j}{\sqrt{\mathrm{num\_heads}}}
  + \mathrm{bias}_{i,j}
\Bigr).
\end{align}

Intuitively, nearby atoms (bonds, angles, dihedrals) should have a higher bias and other atoms should have a lower bias. With a per-DiT-block learnable bias, the DiT has the ability to allocate more attention weight while still retaining the ability to attend globally in some layers.

\subsection{Post-Generation Filtering}

Often, sampling from a flow matching or diffusion model can lead to unphysical generations. In Figure \ref{fig:arch}c, we show post-generation filtering as a sanity check of generation. First, we take the final predicted system $(\text{atom\_type}_i, \hat{b}_z, \hat{f}_i$) and use the Cartesian distances between atoms to calculate the predicted structure, $\hat{\mathcal{G}} = (\mathcal{V}, \hat{\mathcal{E}})$. If the ground truth graph $\mathcal{G}$ and the predicted graph $\hat{\mathcal{G}}$ are isomorphic, i.e., whether $\mathcal{G} \cong \hat{\mathcal{G}}$, then it passes the filter. If any bond length is $<0.8 {\AA}$ then the sample is filtered out. Samples that pass this filtering criteria will be "successful" and those that don't are "unsuccessful." However, a successful prediction of connectivity doesn't guarantee an accurate 3D structure, and the successful samples are evaluated against the optimized 3D structures in Section \ref{sec:results}. 

\subsection{Machine Learning Techniques}

Our main polymer DFT dataset, polyChainStructures (elaborated in Section \ref{sec:dataset}), and the QM9 dataset are imbalanced, with the latter training set being $\approx$ 33x larger. To handle this we upsample the polyChainStructures data by 30x per epoch. During training for the polychain dataset only, this upsampling ratio is held the same for comparison, so that both models see the same amount of polyChainStructures data per epoch. To learn equivariance, we randomly rotate and translate the systems at every training step for both the VAE and the DiT.

The autoencoder models were trained for 300 epochs, and the highest-performing validation checkpoint was taken. The diffusion models were trained for a maximum of 500 epochs.

\section{Data Availability}

The dataset used in this study is available on the Ramprasad group's computational knowledge-base, Khazana (\url{https://khazana.gatech.edu/dataset}).

\section{Code Availability}

The code used for this work will be made publicly available upon publication on the Ramprasad group's github (\url{https://github.com/Ramprasad-Group})

\section{Acknowledgements}

This work was financially supported by the Office of Naval Research (ONR) through Grant N00014-21-1-2258 and the National Science Foundation (NSF) DMREF Grant 2323695. A.J. acknowledges S. Shukla, A. Savit, H. Sahu, H. Tran, and R. Gurnani for valuable discussions.

\section{Author Contributions}

A.J. is the main architect of the models and wrote the paper. R.R. conceived the project and guided the work and the previous project for dataset generation.  

\section{Competing Interests}
R.R. is a founder of Matmerize, Inc., a company specializing in materials informatics software and services. The other authors have no conflicts of interest to declare.



\bibliographystyle{unsrtnat}
{
\small
\bibliography{references}

}

\end{document}